\documentclass[12pt
]{article}
\usepackage[british]{babel}
\usepackage{cite}                      
\usepackage[dvips]{color}              
\usepackage[final]{graphicx}            

\usepackage{multirow}                   
\usepackage{amssymb}
\usepackage{amsmath}
\usepackage{xspace}                    

   \newcommand{\drafty}{}

\newcommand{\non}{\nonumber}
\newcommand{\hq}{\hspace{0.5em}}

\newcommand{\half}{\frac{1}{2}}

\newcommand{\ii}{\mathrm{i}}
\newcommand{\dd}{\mathrm{d}}

\newcommand{\deint}[2]{\dd^{#1}\! #2\;}

\newcommand{\ev}{\vec{e}}
\newcommand{\kv}{\vec{k}}
\newcommand{\pv}{\vec{p}}
\newcommand{\qv}{\vec{q}}

\newcommand{\mpi}{\ensuremath{m_\pi}}
\newcommand{\MeV}{\ensuremath{\mathrm{MeV}}}
\newcommand{\fm}{\ensuremath{\mathrm{fm}}}
\newcommand{\EFTNoPion}{EFT(${\pi\hskip-0.55em /}$)\xspace}

\newcommand{\NXLO}[1]{N\ensuremath{{}^{#1}}LO\xspace}
\newcommand{\NtwoLO}{\NXLO{2}}

\newcommand{\wave}[3]{\ensuremath{{}^{#1}\mathrm{#2}_{#3}}\xspace}
\newcommand{\oneS}{\wave{1}{S}{0}}
\newcommand{\twoS}{\wave{2}{S}{\half}}
\newcommand{\threeS}{\wave{3}{S}{1}}
\newcommand{\fourS}{\wave{4}{S}{\frac{3}{2}}}
\newcommand{\twoPone}{\wave{2}{P}{\half}}
\newcommand{\twoPthree}{\wave{2}{P}{\frac{3}{2}}}
\newcommand{\fourPone}{\wave{4}{P}{\half}}
\newcommand{\fourPthree}{\wave{4}{P}{\frac{3}{2}}}

\newcommand{\ND}{N^\dagger}
\newcommand{\VS}{\vec{\sigma}}

\newcommand{\LRd}{\stackrel{\leftrightarrow}{\de}}
\newcommand{\daR}{g^{(^3 \! S_1-^1 \! P_1)}}
\newcommand{\dbR}{g^{(^1 \! S_0-^3 \! P_0)}_{(\Delta I=0)}}
\newcommand{\dcR}{g^{(^1 \! S_0-^3 \! P_0)}_{(\Delta I=1)}}
\newcommand{\ddR}{g^{(^1 \! S_0-^3 \! P_0)}_{(\Delta I=2)}}
\newcommand{\deR}{g^{(^3 \! S_1-^3 \! P_1)}}

\newcommand{\de}{\partial}
\newcommand{\dev}{\vec{\de}}

\newcommand{\calA}{\mathcal{A}} \newcommand{\calK}{\mathcal{K}}
\newcommand{\calL}{\mathcal{L}} 
\newcommand{\calM}{\mathcal{M}}
\newcommand{\calO}{\mathcal{O}}
\newcommand{\calP}{\mathcal{P}}
\newcommand{\calS}{\mathcal{S}} \newcommand{\calT}{\mathcal{T}}

\newcommand{\mytitle}[1]{\begin{center}\LARGE{\textbf{#1}}\end{center}}
\newcommand{\myauthor}[1]{\textbf{#1}}
\newcommand{\myaddress}[1]{\textit{#1}}
\newcommand{\mypreprint}[1]{\begin{flushright}#1\end{flushright}}

\begin{document}
%

\begin{titlepage}
  \setcounter{page}{0} \mypreprint{
    \drafty
    INT-PUB-10-030\\
    5th July 2010 \\
Revised version 21st July 2010\\
    Accepted by Eur.~Phys.~J.~\textbf{A}
  }
  
  
  \mytitle{On Parity-Violating Three-Nucleon\\ Interactions and the Predictive
    Power\\ of Few-Nucleon EFT at Very Low Energies}


\begin{center}
  \myauthor{Harald W.~Grie\3hammer$^{a,}$\footnote{Corresponding author; email:
      hgrie@gwu.edu}} and \myauthor{Matthias R.~Schindler$^{a}$}\\[2ex]
  
  \vspace*{0.5cm}
  
  \myaddress{$^a$ Center for Nuclear Studies, Department of Physics, \\The
    George Washington University, Washington DC 20052, USA}

  \vspace*{0.2cm}

\end{center}


\begin{abstract}
  We address the typical strengths of hadronic parity-violating three-nucleon
  interactions in ``pion-less'' Effective Field Theory in the nucleon-deuteron
  (iso-doublet) system. By analysing the superficial degree of divergence of
  loop diagrams, we conclude that no such interactions are needed at leading
  order, $\calO(\epsilon Q^{-1})$.  The only two distinct
  parity-violating three-nucleon structures with one derivative mix \twoS and
  \twoPone waves with iso-spin transitions $\Delta I=0$ or $1$.  Due to their
  structure, they cannot absorb any divergence ostensibly appearing at
  next-to-leading order, $\calO(\epsilon Q^0)$. This observation is
  based on the approximate realisation of Wigner's combined $SU(4)$
  spin-isospin symmetry in the two-nucleon system, even when effective-range
  corrections are included.  Parity-violating three-nucleon interactions thus
  only appear beyond next-to-leading order.  This guarantees renormalisability
  of the theory to that order without introducing new, unknown coupling
  constants and allows the direct extraction of parity-violating two-nucleon
  interactions from three-nucleon experiments.
\end{abstract}
\vskip 1.0cm
\noindent
\begin{tabular}{rl}
  Suggested PACS numbers:& \begin{minipage}[t][\height][t]{10.7cm}
    11.10.Ef, 11.10.Gh, 11.30.Er, 11.80.Jy, 13.75.Cs, 21.30.-x, 21.30.Fe,
    21.45.Ff, 25.40.Cm, 25.40.Dn
                    \end{minipage}
                    \\[4ex]
                    Suggested Keywords: &\begin{minipage}[t]{10.7cm}
                      Effective Field Theory, hadronic parity-violation, 
                      three-body system, three-body interaction, 
                      na\"ive dimensional analysis, Wigner's spin-isospin symmetry. 
                    \end{minipage}
\end{tabular}

\vskip 1.0cm

\end{titlepage}

\setcounter{footnote}{0}

\newpage

%

\section{Introduction}
\setcounter{equation}{0}
\label{sec:introduction}

An international effort is under way to map out the weak portion of the
nuclear force at low energies, stimulated in particular by the advancement of
slow-neutron facilities such as the SNS (Oak Ridge), NIST (Gaithersburg), ILL
(Grenoble), PSI (Villingen), FRM-II (M{\"u}nchen).  Experiments on few-nucleon
systems provide an excellent opportunity to study the weak interactions
between free hadrons via parity-violating (PV) effects, see
e.g.~\cite{Adelberger:1985ik,RamseyMusolf:2006dz,Zhu:2004vw}. However, their
interpretation requires adequate theoretical support: The consistency of
different data sets must carefully be checked in one model-independent
framework in order to account for the high complexity of such experiments
which makes their systematic errors difficult to assess; binding effects must
be taken into account with reliable error-estimates; and the PV interaction
strengths must be extracted from PV observables using minimal theoretical
bias, with the long-term goal of relating them to the parameters of the
Standard Model, e.g.~by lattice calculations~\cite{Wasem}.

These criteria are met by calculations in Effective Field Theories (EFTs).
They describe few-nucleon systems with \emph{a priori} estimates of
theoretical uncertainties. Of particular interest to slow-neutron and other
very-low energy facilities is the ``pion-less'' version \EFTNoPion in which
the dynamical degrees of freedom are only nucleons, see
e.g.~\cite{Braaten:2004rn,Platter:2009gz} for recent reviews.  Its range of
applicability is limited to momenta smaller than the pion mass $\mpi$,
i.e.~energies of up to a dozen MeV. This allows for a systematic expansion of
all observables in a generic low-momentum scale $Q$ in units of this breakdown
scale. For typical low momenta like the inverse scattering lengths of the
two-nucleon bound-states, $\gamma\approx45\;\MeV$, the expansion parameter is
usually found to be $Q\approx\frac{1}{[3\dots5]}$. Since external momenta
provide additional scales, the expansion parameter grows when these are
significantly larger than $\gamma$ and finally reaches unity for momenta of
the order of the pion mass. In the parity-conserving (PC) sector, calculations
are routinely performed at next-to-next-to-leading order \NXLO{2} with typical
accuracies of $\lesssim4\%$.  Another expansion parameter
$\epsilon\approx10^{-6}$ is provided by the PV strength relative to the PC
one.  \EFTNoPion was first used in the PV sector in Ref.~\cite{Savage:1998rx},
with a comprehensive description given in Ref.~\cite{Zhu:2004vw}. In the
two-nucleon system, only 5 independent PV parameters exist in the
leading-order Lagrangean, $\calO(\epsilon
Q)$~\cite{Zhu:2004vw,Girlanda:2008ts,Phillips:2008hn}. They have to be
determined from experiment. No further PV 2N interactions enter at NLO,
i.e.~when one power of $Q$ is added. This allows one to compare data, subtract
binding effects and extract PV interactions model-independently with
$\lesssim10\%$ accuracy in $NN$ observables, matching the projected
uncertainties of the most ambitious experiments.

However, the number of feasible low-energy experiments in the PV few-nucleon
sector is limited. A complete data set to determine the PV parameters will
most likely include observables with 3 and more nucleons, where
e.g.~neutron-neutron interactions are probed without the need for free neutron
targets. There are also indications of better PV signals in light nuclei, like
an increased neutron spin rotation in deuterium relative to
hydrogen~\cite{Schiavilla:2008ic}.

The extraction could be thwarted if parity-violating three-nucleon
interactions (3NIs) contribute at the $\gtrsim10\%$-level, i.e.~at LO or NLO
in \EFTNoPion. A PV 3NI will enter at some order as a manifestation of the
interactions underlying \EFTNoPion. Since every PV 3NI requires one additional
experiment to determine its strength, its appearance at LO or NLO would
exacerbate the problem of determining the PV 2NIs from few-nucleon
data. In the strong sector
of \EFTNoPion, a parity-conserving 3NI is expected only at \NXLO{2},
  $\calO(Q^0)$, but already enters at LO, $\calO(Q^{-2})$, because the
anomalously large $NN$ scattering lengths force a non-trivial renormalisation.
Only with a PC 3NI are observables insensitive to physics at short distance
scales.  The PC 3NI strength is determined from one three-nucleon datum,
leading to the Phillips line~\cite{Phillips:1969hm} and Efimov
effect~\cite{efimov}; see e.g.~\cite{Braaten:2004rn,Platter:2009gz} for
reviews. Such a non-perturbative renormalisation may interfere with PV 3NIs
and promote them to lower orders than simplistically expected.

We show that no PV 3NI enters at LO ($\calO(\epsilon Q^{-1})$) or NLO ($\calO(\epsilon Q^{0})$) in
the nucleon-deuteron system, the only 3N system which can be tested
experimentally. We draw from Ref.~\cite{effrange1,effrange2}, where the
general method to determine the divergence structure of 3N amplitudes was
presented.  Section~\ref{sec:Lags} briefly recalls those aspects of PC and PV
\EFTNoPion needed here. We then proceed in two steps: Na\"ive dimensional
analysis in Sec.~\ref{sec:LOamplitudes} shows that there are no PV 3NIs at
leading order.  We then construct all PV 3NIs with only one derivative in
Sec.~\ref{sec:PV3NI}.  Section \ref{sec:NLOgraphs} discusses which PV
nucleon-deuteron scattering contributions arise at NLO. In
Sec.~\ref{sec:NLOamplitudes}, we show that the constructed PV 3NIs do not
match the structure of possible divergences, so that no PV 3NIs exist even at
NLO. A summary with potential limitations and extensions of this work
concludes the article.

\section{Interactions and UV Limit}
\setcounter{equation}{0}
\label{sec:Lags}

\subsection{Parity-Conserving Part}
\label{sec:pclag}

In the parity-conserving three-nucleon sector of \EFTNoPion, we follow the
conventions of Ref.~\cite{Griesshammer:2004pe}. The pertinent
pieces of the parity-conserving Lagrangean up to NLO are:
\begin{align}\label{eq:PCLag}
  \mathcal{L}_\text{PC} =& \ND(i \de_0 + \frac{\vec{\de}^2}{2M})N -y\left[ d_t^{i
      \dagger} (N^T P^i_t N) +\mathrm{H.c.}\right]
  -y\left[ d_s^{A \dagger} (N^T P^A_s N) +\mathrm{H.c.}\right]\\
  & +d_t^{i\dagger}\left[\Delta_t
    -c_{0t}\left(i\partial_0+\frac{\dev^2}{4M}+\frac{\gamma_t^2}{M}\right)
  \right] d_t^i +d_s^{A\dagger}\left[\Delta_s
-c_{0s}\left(i\partial_0+\frac{\dev^2}{4M}+\frac{\gamma_s^2}{M}\right)\right]
  d_s^A\notag\\&
  +\frac{y^2M\,H_0(\Lambda)}{3\Lambda^2}
  \left[d^i_t(\sigma_iN)-d_s^A(\tau_AN)\right]^\dagger
  \left[d^i_t(\sigma_iN)-d_s^A(\tau_AN)\right]+\ldots \notag
\end{align} 
The nucleon field $N$ has mass $M$. The spin-triplet and spin-singlet dibaryon
fields $d_t$ and $d_s$ are introduced as auxiliary fields with the quantum
numbers of the corresponding S-wave two-nucleon states to simplify
calculations~\cite{Kaplan:1996nv,Bedaque:1999vb}. With $\sigma_i$ ($\tau_A$)
denoting Pauli matrices in spin (iso-spin) space, $P_t^i=\frac{1}{\sqrt{8}}
\tau_2 \sigma_2 \sigma_i$ and $P_s^A=\frac{1}{\sqrt{8}}\tau_2\tau_A \sigma_2$
project the two-nucleon state onto the \threeS and \oneS partial waves (in the
notation $^{2S+1}{l}_J$, with $S$ the spin, $l$ the orbital angular momentum
and $J$ the total angular momentum). The parity-conserving \twoS-wave
three-nucleon interaction in the last line has strength $H_0(\Lambda)$ which
depends on a regulator $\Lambda$, see below. We choose $y^2=4\pi/M\sim Q^0$.
The LO parameters $\Delta_{s/t}$ are determined from low-energy data, e.g.~the
poles of the $NN$ $S$-wave amplitudes at $\ii\gamma_{s/t}$, and are the only
terms of unnatural size, $\Delta_{s/t}\sim Q^{-1}$. The auxiliary-field
propagators are at leading order:
\begin{equation}
  \label{eq:dprop} 
  D_{s/t}(q_0,\qv)=\frac{1}{\gamma_{s/t}-
  \sqrt{\frac{\qv^2}{4}-Mq_0-\ii\epsilon}}\;\;.
\end{equation}
The parameters $c_{0s/t}$ enter at NLO, determined
for example by the effective ranges~\cite{Phillips:1999hh,Griesshammer:2004pe}.

\begin{figure}[!hbt]
\begin{center}
     \includegraphics*{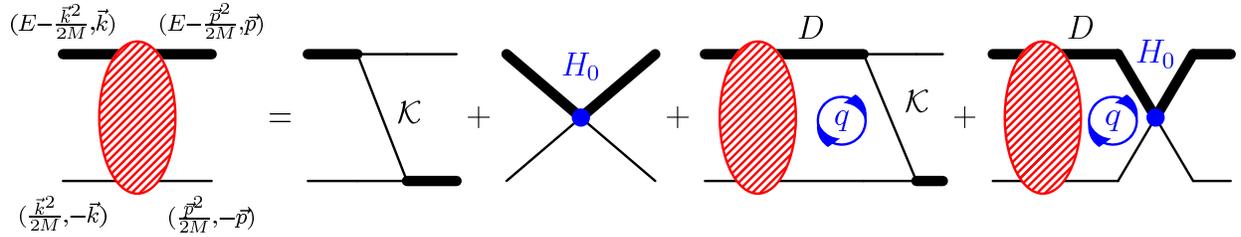}
     \caption{Three-nucleon scattering equation.  Thick line: two-nucleon
       auxiliary-field propagator $D_{s/t}$ (\oneS/\threeS); thin line
       ($\calK$): propagator of the exchanged nucleon; $H_0$: PC 3NI.}
\label{fig:3NAmpt}
\end{center}
\end{figure}

The Faddeev equations for the half-offshell amplitudes of nucleon-deuteron
scattering at LO in the $l$th partial wave before wave-function
renormalisation were first derived by Skorniakov and
Ter-Martirosian~\cite{skorny}. Generalisation to full-offshell amplitudes is
straight-forward. The kinematics in the centre-of-mass system is specified in
Fig.~\ref{fig:3NAmpt}, with total non-relativistic energy $E$; momentum $\kv$
for the incoming deuteron; momentum $\pv$ for the outgoing one. The amplitude
for half-offshell momenta $p=|\pv|$ is found by setting the incoming leg on-shell,
$E=\frac{3\kv^2}{4M}-\frac{\gamma_t^2}{M}$; the on-shell point is in addition
at $p=k$.
The propagator of the exchanged nucleon, projected onto orbital angular
momentum $l$, is
\begin{equation}
  \label{eq:projectedNpropagator}
  \calK^{(l)}(E;q,p):=\frac{1}{2}\;\int\limits_{-1}^1\deint{}{\cos\theta}
  \frac{P_l(\cos\theta)}{p^2+q^2-ME+ pq\cos\theta}
  =\frac{(-1)^l}{pq}\;Q_l\left(\frac{p^2+q^2-ME}{pq}\right),
\end{equation}
with $\theta=\angle(\pv;\qv)$ and $P_l(z)$ ($Q_l(z)$) the $l$th Legendre
polynomial of the first (second) kind with complex argument~\cite{Gradshteyn}.

Two spin channels exist in the 3N system. The total spin $S=\frac{3}{2}$
(quartet) channel only receives contributions from combining the nucleon with
the spin-triplet auxiliary field $d_t$, while both $d_t$ and the spin-singlet
auxiliary field $d_s$ contribute in the $S=\frac{1}{2}$ (doublet) channel.
With the two configurations $d_t N$ and $d_s N$, it is convenient to follow
Ref.~\cite[App.~A.1]{effrange1,effrange2} in representing operators $\calO$ by
a $2\times2$-matrix in the so-called cluster-decomposition space:
\begin{equation}
  \calO = N^\dagger_{b\beta} 
  \;\left(d^\dagger_{t,j},\;d^\dagger_{s,B}\right)
  \begin{pmatrix}\calO(Nd_t\to Nd_t)^j_i&\calO(Nd_s\to Nd_t)^j_A\\
    \calO(Nd_t\to Nd_s)^B_i&\calO(Nd_s\to Nd_s)^B_A
  \end{pmatrix}^{b\beta}_{\hq\hq a\alpha} 
  \;{d_t^i\choose d_s^A}
  \;N^{a\alpha}
  \;\;.
\label{eq:clusterdecomp}
\end{equation}
Operators thus act in the direct tensor-product space
$\textbf{spin}\otimes\textbf{iso-spin}\otimes\textbf{cluster}$ and carry the
following indices: vector $i,j$, iso-vector $A,B$, spinor $ \alpha,\beta$ and
iso-spinor $a,b$. The latter two will often be suppressed for convenience. The
LO offshell amplitude
$t^{(l)}_q(E;k,p)\left(\begin{smallmatrix}1&0\\0&0\end{smallmatrix}\right)$ in
the spin-quartet channels is the solution to the integral equation
\begin{equation}
  \label{eq:faddeev}
  {t^{(l)}_q
    }(E;k,p)=
    -4\pi\; {\calK^{(l)}(E;k,p)
    }
    +\frac{2}{\pi}\;
    \int\limits_0^\Lambda\deint{}{q} q^2\;\calK^{(l)}(E;q,p)\;
    D_t(E-\frac{q^2}{2M},q)\;{t^{(l)}_q
    }(E;k,q)\;\;.
\end{equation}
The regulator $\Lambda$ is used in the following to study the UV limit of
Eq.~(\ref{eq:faddeev}). In cutoff regularisation, $\Lambda$ is equal to or
larger than the scale at which \EFTNoPion breaks down.  

For $S=\half$, the amplitude $t^{(l)}_{d,XY}$ stands for the $Nd_X\to
Nd_Y$-process, where $X,Y=s$ or $t$. For example, $t^{(l)}_{d,ts}$ stands for
$Nd_t\to Nd_s$. Following e.g.~Ref.~\cite[App.~A]{Griesshammer:2004pe}, the
full-offshell amplitude is the $2\times2$ matrix which solves
\begin{eqnarray}
 \lefteqn{
\begin{pmatrix}t^{(l)}_{d,tt}&t^{(l)}_{d,st}\\t^{(l)}_{d,ts}&t^{(l)}_{d,ss}\end{pmatrix}
  (E;k,p)=2\pi\;
  \left[\calK^{(l)}(E;k,p)\;\begin{pmatrix}1&-3\\-3&1\end{pmatrix}
    +\delta^l_0\;\frac{2H_0(\Lambda)}{\Lambda^2}\;
  \begin{pmatrix}1&-1\\-1&1\end{pmatrix}\right]}\non\\
  \label{eq:doubletpw}
  &&-\;\frac{1}{\pi}\int\limits_0^\Lambda\deint{}{q} q^2\;\left[
    \calK^{(l)}(E;q,p)\;\begin{pmatrix}1&-3\\-3&1\end{pmatrix}
    +\delta^{l0}\;\frac{2H_0(\Lambda)}{\Lambda^2}\;\begin{pmatrix}1&-1\\-1&1\end{pmatrix}
  \right]\;
  \\&&
  \hq\hq\hq\hq\hq\hq\hq\hq\hq\hq
  \times\;\begin{pmatrix}D_t&0\\0&D_s\end{pmatrix}(E-\frac{q^2}{2M},q)
  \;\begin{pmatrix}t^{(l)}_{d,tt}&t^{(l)}_{d,st}\\t^{(l)}_{d,ts}&t^{(l)}_{d,ss}
  \end{pmatrix}
  (E;k,q)\;\;,
  \non
\end{eqnarray}
and the half-offshell amplitude with an incoming nucleon and deuteron is
obtained by multiplying with the column vector ${1\choose0}$ from the right
and setting $E=\frac{3\kv^2}{4M}-\frac{\gamma_t^2}{M}$.

According to a simplistic dimensional estimate, the three-nucleon interaction
$H_0$ is of higher order and should thus not be included in
Eq.~(\ref{eq:doubletpw}). However, a detailed analysis of $nd$ scattering in
the \twoS-wave~\cite{Bedaque:1999ve} shows significant dependence on $\Lambda$
in the solution of the three-body equation without $H_0$. This
cutoff-dependence is removed by promoting the 3NI to leading order.  Its
strength $H_0$ is determined from one three-nucleon datum like the triton
binding energy. The Phillips line~\cite{Phillips:1969hm} and Efimov
effect~\cite{efimov} emerge as the physically observable remnants of the UV
limit-cycle which describes its renormalisation-group running, see
e.g.~\cite{Braaten:2004rn,Platter:2009gz} for reviews.

In order to investigate whether a similar promotion of higher-order terms
occurs for PV 3NIs in \EFTNoPion, we must consider the UV-limit of the
half-offshell momenta of (\ref{eq:faddeev}/\ref{eq:doubletpw}). For $p,q\gg
\sqrt{ME},\,k,\;\gamma_{s/t}$, the auxiliary-field propagators are independent of
external scales,
\begin{equation}
  \label{eq:dforWigner}
  \lim\limits_{q\gg \sqrt{ME},\,\gamma_{s/t}}D_{s/t}(E-\frac{\qv^2}{2M},\qv)=
  -\frac{2}{\sqrt{3}} \;\frac{1}{q}\;\;,
\end{equation} 
as is the kernel $\calK$. It has been demonstrated before~\cite{effrange1,effrange2,GM}
that the solutions to the resulting integral equations in the UV limit are
linear combinations of
\begin{equation}
  \label{eq:solution}
  t^{(l)}_\lambda(q):=\lim\limits_{q\gg \sqrt{ME},\,
    k,\;\gamma_{s/t}}t^{(l)}_\lambda(E;k,q) 
  \propto k^l\;q^{-s_l(\lambda)-1}\;\;,
\end{equation}
with the asymptotic exponents $s_l(\lambda)$ of the amplitudes at large
half-offshell momenta $p,q\gg \sqrt{ME},k,\gamma_{s/t}$ given for the lowest
angular momenta in Table~\ref{tab:svalues}.
\begin{table}[!ht]
  \centering
  \begin{tabular}{|c||l|l|}
    \hline
    \rule[-1.5ex]{0ex}{4ex}
    partial wave $l$&$s_l(\lambda=1)$&$s_l(\lambda=-\half)$\\
    \hline
    \hline
    \rule[-1.5ex]{0ex}{4ex}
    0&$\pm1.00624\dots\;\ii$&2.16622\dots\\
    \hline
    \rule[-1.5ex]{0ex}{4ex}
    1&2.86380\dots&1.77272\dots\\
    \hline
    \rule[-1.5ex]{0ex}{4ex}
    $l\ge2$&$\approx l+1$&$\approx l+1$\\
    \hline
  \end{tabular}
  \caption{Asymptotic coefficients $s_l(\lambda)$.}
  \label{tab:svalues}
\end{table}
For the spin-quartet channels, the spin-isospin parameter is $\lambda=-\half$.
For the spin-doublet channels, the situation is slightly more complicated. In
the UV limit, the two auxiliary-field propagators \eqref{eq:dprop} are
identical and $NN$ scattering becomes automatically Wigner-$SU(4)$-symmetric,
i.e.~symmetric under arbitrary combined rotations of spin and
iso-spin~\cite{Wigner:1936dx,Bedaque:1999ve,Mehen:1999qs}. The integral
equations can then be decoupled by the transformation
\begin{equation}
  \label{eq:lincombforWigner}
  \begin{pmatrix}t^{(l)}_{\lambda=1}&t^{(l)}_{\lambda=-\half\to\lambda^\prime=1}\\
    t^{(l)}_{\lambda=1\to\lambda^\prime=-\half}&t^{(l)}_{\lambda=-\half}\end{pmatrix}
  =\half\begin{pmatrix}1&-1\\1&1\end{pmatrix}\;
  \begin{pmatrix}t^{(l)}_{d,tt}&t^{(l)}_{d,st}\\t^{(l)}_{d,ts}&t^{(l)}_{d,ss}\end{pmatrix}   \begin{pmatrix}1&1\\-1&1\end{pmatrix}
  \;\;,
\end{equation}
leading to the Faddeev equation in the ``Wigner-basis'',
\begin{eqnarray}
 \lefteqn{
\begin{pmatrix}t^{(l)}_{1}&t^{(l)}_{-\half\to1}\\
  t^{(l)}_{1\to-\half}&t^{(l)}_{-\half}\end{pmatrix}
  (E;k,p)=2\pi\;
  \left[\calK^{(l)}(E;k,p)\;\begin{pmatrix}2&0\\0&-1\end{pmatrix}
    +\delta^l_0\;\frac{2H_0(\Lambda)}{\Lambda^2}\;
  \begin{pmatrix}1&0\\0&0\end{pmatrix}\right]}\non\\
  \label{eq:doubletpwWigner}
  &&-\;\frac{2}{\pi}\int\limits_0^\Lambda\deint{}{q} q^2\;\left[
    \calK^{(l)}(E;q,p)\;\begin{pmatrix}2&0\\0&-1\end{pmatrix}
    +\delta^{l0}\;\frac{2H_0(\Lambda)}{\Lambda^2}\;\begin{pmatrix}1&0\\0&0\end{pmatrix}
  \right]\;
  \\&&
  \hq\hq\hq\hq\hq\hq\hq\hq\hq\hq
  \times\;\begin{pmatrix}\Sigma&\Delta\\\Delta&\Sigma\end{pmatrix}(E-\frac{q^2}{2M},q)
  \begin{pmatrix}t^{(l)}_{1}&t^{(l)}_{-\half\to1}\\
  t^{(l)}_{1\to-\half}&t^{(l)}_{-\half}\end{pmatrix}
  (E;k,q)\;\;,\non
\end{eqnarray}
in which only the auxiliary-field propagators are not diagonal. While
$\Sigma=\half(D_t+D_s)$ is the ``average'' $NN$ S-wave amplitude,
$\Delta=\half(D_t-D_s)$ parameterises the degree to which the \threeS and
\oneS-amplitudes differ~\cite{Bedaque:2002yg,Griesshammer:2004pe}. In the
UV-limit, $\Delta=0$ from \eqref{eq:dforWigner}, the components decouple in
the Wigner-basis, and full-offshell amplitudes are diagonal. Off-diagonal
elements are suppressed by $(\gamma_t-\gamma_s)/q$ in the UV limit and hence
by one power of $(\gamma_s-\gamma_t)/\mpi\sim
Q$~\cite{Hammer:2001gh,Bedaque:2002yg,Griesshammer:2004pe}. The eigenvectors
$t_1^{(l)}$ and $t_{-\half}^{(l)}$ are also half-offshell amplitudes when
$E=\frac{3\kv^2}{4M}-\frac{\gamma_t^2}{M}$, see
e.g.~\cite{Griesshammer:2004pe}.

The amplitude $t_{1}^{(l)}$ is the eigenvector to the spin-isospin
parameter $\lambda=1$ and obeys the same integral equation as three
spinless bosons; the amplitude $t_{-\half}^{(l)}$ is the
  eigenvector to $\lambda=-\half$ and shows the same asymptotics as the
spin-quartet amplitudes Eq.~\eqref{eq:faddeev}.  Since the asymptotic
coefficient $s_0(\lambda=1)=\pm1.0062\dots\ii$ in the \twoS-wave is imaginary,
the two solutions are super-imposed~\cite{Braaten:2004rn,Platter:2009gz}:
\begin{equation}
  \label{eq:efimov}
  t^{(l=0)}_{1}(q)\propto\frac{\cos[1.0062\dots\ln[q]+\delta]}{q}\;.
\end{equation}
The relative phase $\delta$ is related to the PC 3NI strength $H_0(\Lambda)$
and thus determined by one 3N datum in the PC sector.

\begin{figure}[!htb]
  \begin{center}
    \includegraphics*{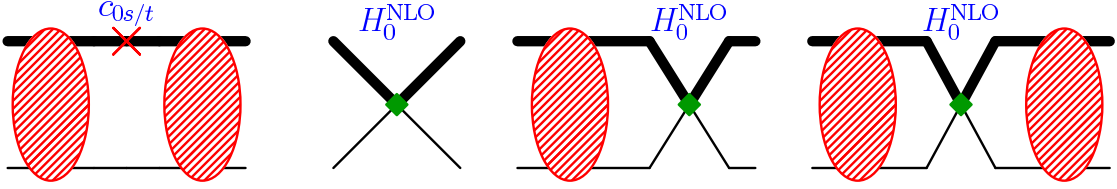}
    \caption{NLO corrections to the LO PC amplitudes: effective-range (cross);
      PC 3NI $H_0^\text{NLO}$ (diamond). Crossed contributions not displayed.}
    \label{fig:PCNLO}
  \end{center}
\end{figure}

At NLO, exactly two corrections enter, as pictured in Fig.~\ref{fig:PCNLO}.
Both are central, so that partial-waves do not mix.
The first interaction is the insertion of one effective-range term $c_{0s/t}$
of the PC Lagrangean \eqref{eq:PCLag} in an auxiliary-field propagator. It is
diagonal in the partial-wave basis since the two auxiliary fields $d_s$ and
$d_t$ do not mix, cf.~Eq.~\eqref{eq:PCLag}. In the Wigner-basis of
Eq.~\eqref{eq:lincombforWigner}, this translates to
\begin{equation}
  \label{eq:effrangewigner}
  \half\begin{pmatrix}1&-1\\1&1\end{pmatrix}  \begin{pmatrix}c_{0t}&0\\0&c_{0s}\end{pmatrix}
  \begin{pmatrix}1&1\\-1&1\end{pmatrix}=
  \half\begin{pmatrix}c_{0t}+c_{0s}&c_{0t}-c_{0s}\\
    c_{0t}-c_{0s}&c_{0t}+c_{0s}\end{pmatrix}\;\;.
\end{equation}
When $c_{0s/t}=\rho_{0s/t}M/2$ are determined by the effective ranges
$\rho_{0s/t}$~\cite{Phillips:1999hh,Griesshammer:2004pe} with physical values
$\rho_{0s}=2.73\,\fm$ and $\rho_{0t}=1.76\,\fm$, one finds that the
off-diagonal elements are suppressed by a factor $Q$ relative to the diagonal
ones:
\begin{equation}
  \frac{c_{0t}-c_{0s}}{c_{0t}+c_{0s}}\approx-0.22\sim Q\;.
\end{equation}
Off-diagonal elements again only contribute  at higher orders in the
power-counting:
\begin{equation}
\half\begin{pmatrix}c_{0t}+c_{0s}&0\\0&c_{0t}+c_{0s}\end{pmatrix} + \calO(Q).
\end{equation}
Determining $c_{0s/t}$ by different low-energy data leads to results which
only differ by higher orders in $Q$. Off-diagonal elements are particularly
strongly suppressed in
Z-parameterisation~\cite{Phillips:1999hh,Griesshammer:2004pe}, where $c_0$ is
determined by the residue of the pole in the $NN$ amplitude as
$c_{0s/t}=\frac{M\rho_{0s/t}}{2(1-\gamma_{s/t}\rho_{0s/t})}$, so that
$\frac{c_{0t}-c_{0s}}{c_{0t}+c_{0s}}\approx\frac{1}{10}\sim Q^2$. The $NN$
scattering amplitudes in the \oneS and \threeS channels therefore are to a
good approximation Wigner-symmetric even when the effective-range corrections
are taken into account.

The second NLO correction comes from including the NLO piece of the
momentum-independent PC 3NI in the \twoS-wave, $H_0^\text{NLO}(\Lambda)$. It
must be inserted once and re-adjusted to renormalise the NLO PC
amplitudes~\cite{Hammer:2001gh,Bedaque:2002yg}. The NLO parity-conserving 3NI
$H_0^\text{NLO}$ of Eq.~\eqref{eq:PCLag} in the Wigner-basis can be read off
from Eq.~\eqref{eq:doubletpwWigner} as proportional to
$\left(\begin{smallmatrix}1&0\\0&0\end{smallmatrix}\right)$.
Corrections are again suppressed by one power of Q and hence are of higher
order~\cite{Hammer:2001gh,Bedaque:2002yg,Griesshammer:2004pe}. 

That the LO 3N amplitude and its NLO corrections are diagonal in the
Wigner-basis in the UV limit implies that in the parity-conserving sector, the
spin-isospin parameter $\lambda$ is approximately a good quantum number in the
Wigner-SU$(4)$ limit. This will be fundamental for showing in
Sec.~\ref{sec:NLOamplitudes} that no PV 3NIs exist at NLO.

\subsection{Parity-Violating Two-Nucleon Part}
\label{sec:pvlag}

In contradistinction to the parity-conserving sector, parity-violating
interactions may be included perturbatively at leading order since they are
suppressed by $\epsilon\sim10^{-6}$. The parity-violating two-nucleon
Lagrangean at leading order, $\calO(\epsilon Q)$, in the dibaryon formalism
contains 5 coupling constants describing mixtures between S- and P-waves with
the same total angular momentum and different iso-spin transitions, see
Ref.~\cite{Girlanda:2008ts,Phillips:2008hn,Schindler:2009wd} for details:
\begin{align}\label{eq:PVLag}
  \mathcal{L}_\text{PV}^\text{LO}= - & \left[ \daR d_t^{i\dagger} \left(N^T
      \sigma_2 \tau_2\,\ii\!\!\LRd_i N\right) \right. \notag\\
  & +\dbR d_s^{A\dagger}
  \left(N^T\sigma_2 \ \VS \cdot \tau_2 \tau_A \,\ii\!\!\LRd  N\right) \notag\\
  & +\dcR \ \epsilon^{3AB} \, d_s^{A\dagger}
  \left(N^T \sigma_2  \ \VS\cdot \tau_2 \tau^B \LRd N\right) \notag\\
  & +\ddR \ \mathcal{I}^{AB} \, d_s^{A\dagger}
  \left(N^T \sigma_2 \ \VS\cdot \tau_2 \tau^B \,\ii \!\!\LRd N\right) \notag\\
  & \left. +\deR \ \epsilon^{ijk} \, d_t^{i\dagger} \left(N^T \sigma_2
      \sigma^k \tau_2 \tau_3 \LRd{}^{\!j} N\right) \right] 
  +\mathrm{H.c.}
\end{align}
Here, $\LRd=\stackrel{\rightarrow}{\de}-\stackrel{\leftarrow}{\de}$,
$\mathcal{I}=\mathrm{diag}(1,1,-2)$ is a diagonal matrix in iso-vector space,
and we have omitted couplings to external currents. The resulting
leading-order PV two-nucleon scattering amplitude counts as $\calO(\epsilon
Q^{0})$ because $NN$ rescattering is enhanced by factors of $\Delta\sim
Q^{-1}$ in the auxiliary-field formalism, while all other parameters are of
natural size~\cite{Griesshammer:2004pe}. Any new interaction for S-P or S-F
transitions must contain at least three derivatives and hence enters at
\NXLO{2}, $\calO(\epsilon Q^3)$, which makes them relevant only for
calculations with $\lesssim3\%$ accuracy.

The three-nucleon amplitudes containing PV two-nucleon interactions are
obtained from \eqref{eq:PVLag}. But since the Effective Field Theory paradigm
is that the Lagrangean contains \emph{all} interactions allowed by symmetries,
the question arises at which order the first PV 3NI appears. The lowest-order
contributions will again connect S and P waves, analogous to the two-nucleon
sector.

\section{Parity-Violating Three-Nucleon Interactions in the Nucleon-Deuteron
System}
\setcounter{equation}{0}
\label{sec:amplitudes}

\begin{figure}[!htb]
\begin{center}
     \includegraphics*{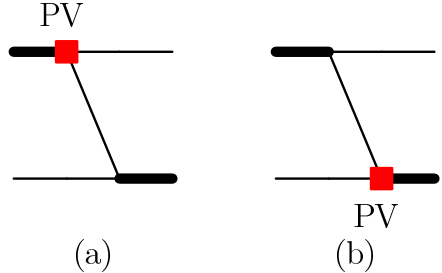}
     \caption{LO tree-level PV diagrams. Square (PV): the parity-violating 2N
       vertices $S\leftrightarrow P$.}
\label{fig:tree}
\end{center}
\end{figure}

Simplistically, one may attempt to derive the order of the first PV 3NI by
considering the tree-level diagrams in Fig.~\ref{fig:tree}.
The PV 2NI scales as $\epsilon Q$ since it contains one derivative, see
Lagrangean \eqref{eq:PVLag}. As the nonrelativistic propagator of the
exchanged nucleon \eqref{eq:projectedNpropagator} scales as $Q^{-2}$, the
tree-level diagrams count as $\calO(\epsilon Q^{-1})$.  The first nonzero PV
3NI mixes S- and P-waves and thus contains one derivative, which translates
into one power of $Q$.  The PV 3NI therefore seems to enter at order $\epsilon
Q$ or \NtwoLO i.e.~suppressed by 2 powers of $Q$ against the first tree-level
diagram involving PV 2NIs.  In other channels, like P-D and S-F, a PV 3NI
includes additional derivatives and hence additional powers of $Q$,
suppressing their contribution even further.  We therefore consider only PV
3NIs in the four channels of the $Nd$ system which mix S- and P-waves and
conserve total angular momentum:
\begin{equation}
  \label{eq:channels}
  \twoS - \twoPone\;,\; \twoS - \fourPone\;\;;\;\;
  \fourS - \twoPthree\;,\;\fourS - \fourPthree
\end{equation}
As discussed above, large scattering lengths in the two-nucleon system lead to
non-trivial renormalisation in the PC sector of both the two- and
three-nucleon system, which in turn results in the promotion of, for example,
the \twoS PC 3NI to lower order than the simplistic argument predicted. A more
careful analysis in the PV sector is therefore warranted in particular in
amplitudes which involve the critical \twoS-wave in the initial or final
state.

\subsection{No PV Three-Nucleon Interaction at Leading Order}
\label{sec:LOamplitudes}

\begin{figure}[!htb]
  \begin{center}
    \includegraphics*{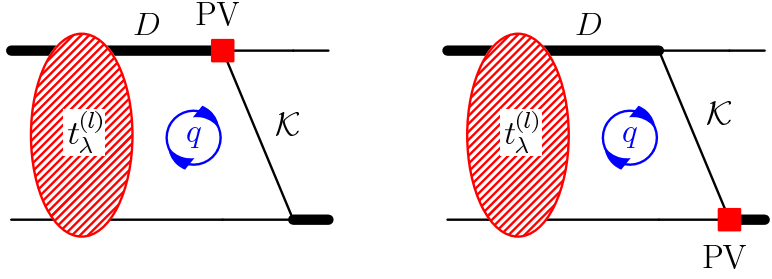}
    \caption{LO one-loop PV diagrams. Crossed contributions not displayed.}
    \label{fig:oneloop}
  \end{center}
\end{figure}

Consider the divergences generated by PV 2NIs in irreducible 3N diagrams. No
loops and thus no divergences exist at tree level, Fig.~\ref{fig:tree}.
One-loop diagrams, Fig.~\ref{fig:oneloop}, contain the LO PC 3N half-offshell
amplitude $t^{(l)}_{\lambda}(E;k,q)$ once, with $q$ the loop momentum.
The LO PV 2N vertex itself contains a momentum-dependent piece proportional to
loop and outgoing momenta. We symbolically denote the PV nucleon exchange as
$(\pv\text{ or }\qv)\cdot\epsilon\vec{K}_{PV}$, where $\epsilon\vec{K}_{PV}$
is a stand-in for PV and PC couplings and spin-isospin-cluster structure, but
does not contain the vectors $\qv$ or $\pv\sim\kv$, or their magnitudes. Since
it contributes one unit of angular momentum and violates parity, the
transition amplitude relates states with orbital angular momenta $l$ and
$l\pm1$. Performing the energy integration by picking up the spectator nucleon
pole at $q_0=\frac{q^2}{2M}$, the transition amplitude is made up of terms
with the symbolic form
\begin{equation}
  \label{eq:oneloop}
  \int\limits^\Lambda\deint{}{q}q^2\int\deint{}{\cos\theta} 
  \frac{(\pv\text{ or }\qv)\cdot\epsilon\vec{K}_{PV}}{p^2+q^2-ME+ \pv\cdot\qv}
  \;P_{l}(\cos\theta)\;\frac{1}{\gamma_{s/t}-\sqrt{\frac{3q^2}{4}-ME}}\;
  t^{(l)}_{\lambda}(E;k,q)
\end{equation}
for each component in cluster-space. As in \eqref{eq:projectedNpropagator},
$\int\deint{}{\cos\theta}$ projects the exchange-term to match the orbital
angular momentum of the incoming PC amplitude $t^{(l)}_\lambda$.

The UV limit $q\gg \sqrt{ME},p,k,\gamma_{s/t}$ of the amplitude is now constructed
from that of its constituents: Either of the auxiliary-field propagators in
the \threeS and \oneS state approaches $1/q$, see \eqref{eq:dforWigner}; the
asymptotics of $t_\lambda^{(l)}$ is given by the spin-isospin-dependent
exponent $s_l(\lambda)$ of \eqref{eq:solution} with Table~\ref{tab:svalues};
and only the intermediate-nucleon propagator and the PV vertex depend on
$\theta$, with the propagator expanded in powers of $\frac{p}{q}$. The
asymptotics is therefore
\begin{equation}
  \label{eq:oneloopasymptotics}
  \int\limits^\Lambda\frac{\deint{}{q}}{q^{2+s_l(\lambda)}}\int\deint{}{\cos\theta}
  P_l(\cos\theta)\;(\pv\text{ or }\qv)\cdot\epsilon\vec{K}_{PV}
  \left[1-\frac{\pv\cdot\qv-\frac{2}{\sqrt{3}}\,q\gamma_{s/t}}{q^2}
    +\calO(q^{-2})\right]\;\;,
\end{equation}
where $\calO(q^{-2})$ denotes further corrections to the auxiliary-field and
intermediate-nucleon propagators suppressed by two powers of
$p,\,k,\,\gamma_{s/t}$ over $q$.

Combining the first term in brackets with the most divergent piece
$\qv\cdot\vec{K}_{PV}$ of the interaction, the most UV-dependent contribution
for a given orbital angular momentum $l$ of the PC amplitude thus seems to
scale as $q^{-s_l(\lambda)}$, i.e.~its superficial degree of divergence seems
to be $\Delta(l;\lambda)=-\textrm{Re}[s_l(\lambda)]$. Inspecting
Table~\ref{tab:svalues}, one finds the diagram converges for nearly all
partial waves, except when the PC amplitude is the $\lambda=1$ part of the
\twoS-wave ($\lambda=1$ and $l=0$). Since $s_0(\lambda=1)=1.006\dots\ii$ is
imaginary, the PV amplitude appears logarithmically divergent, with a phase
determined by a PC three-nucleon datum, see Eq.~\eqref{eq:efimov}. However,
this amplitude is actually identically zero upon angular integration:
$\int\deint{}{\cos\theta}P_0(\cos\theta)\qv=0$. The next term in the bracket,
$\pv\cdot\qv/q^2$, is nonzero after angular integration,
$\int\deint{}{\cos\theta}P_0(\cos\theta)\qv(\pv\cdot\qv)/q^2\propto \pv$, but
suppressed by one more inverse power of $q$. This part converges thus with
degree $\Delta=-1-\textrm{Re}[s_0(\lambda)]<0$.  For the other part of the
interaction, $\pv\cdot\vec{K}_{PV}$, the angular integration from the
``$1$''-term of the bracket is nonzero. The degree of divergence is again
given by $\Delta=-1-\textrm{Re}[s_0(\lambda)]<0$, and the PV S-to-P wave
amplitude therefore scales as:
\begin{equation}
  \label{eq:LOdivergenceS}
  \lim\limits_{q\to\infty}\calA^{(\text{S}\to\text{P})}_{LO,1\text{-loop}}\sim
  (\pv\cdot\vec{K}_{PV})\;q^{-1-s_0(\lambda)}\to0
  \;\;,\;\;\Delta^{(\text{S}\to\text{P})}_{LO,1\text{-loop}}(\lambda)=
  -1-\textrm{Re}[s_0(\lambda)]
\end{equation}
Further terms in the expansion are suppressed by more negative powers of %
$q$. From Table~\ref{tab:svalues}, one reads off
$s_0(\lambda=1)=1.006\dots\ii,\;s_0(\lambda=-\half)=2.166\dots$ and concludes
that no divergences occur when LO PV 2NIs are convoluted with LO PC one-loop
S-wave amplitudes, since the partial-wave-dependent degrees of divergence are:
\begin{equation}
  \label{eq:LOdegreeS}
  \Delta^{(\text{S}\to\text{P})}_{LO,1\text{-loop}}(\lambda=1)=-1
  \;\;,\;\;
  \Delta^{(\text{S}\to\text{P})}_{LO,1\text{-loop}}(\lambda=-\half)=-3.16\dots
\end{equation}
When convoluting with a PC P-wave amplitude, the angular integral involving
the ``$1$''-term is $ \int\deint{}{\cos\theta}P_1(\cos\theta)\qv\not=0$. The
amplitude thus scales as:
\begin{equation}
  \label{eq:LOdivergenceP}
  \lim\limits_{q\to\infty}\calA^{(\text{P}\to\text{S})}_{LO,1\text{-loop}}\sim
  (\pv\cdot\vec{K}_{PV})\;q^{-s_1(\lambda)}\to0
  \;\;,\;\;\Delta^{(\text{P}\to\text{S})}_{LO,1\text{-loop}}(\lambda)= 
  -\textrm{Re}[s_1(\lambda)]
\end{equation}
and, using Table~\ref{tab:svalues}, the superficial degrees of divergence are
hence:
\begin{equation}
  \label{eq:LOdegreeP}
\Delta^{(\text{P}\to\text{S})}_{LO,1\text{-loop}}(\lambda=1)=-2.86\dots
\;\;,\;\;
\Delta^{(\text{P}\to\text{S})}_{LO,1\text{-loop}}(\lambda=-\half)=-1.77\dots
\end{equation}
i.e.~no divergence occurs. By time-reversal symmetry, LO one-loop
contributions with the PC rescattering amplitude on the outgoing leg share the
same divergence structure.

\begin{figure}[!htb]
  \begin{center}
    \includegraphics*{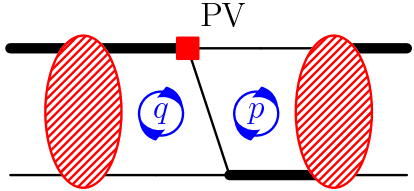}
    \caption{LO two-loop PV diagram. Crossed contributions not displayed.}
    \label{fig:twoloop}
  \end{center}
\end{figure}

Consider now two-loop amplitudes, Fig.~\ref{fig:twoloop}. The same argument
applies separately for each integration when the loop momenta are $q\gg p$ or
$p\gg q$.
The overlapping divergence for $p\sim q\gg \sqrt{ME},k,\gamma_{s/t}$ leads to
overall scaling
\begin{equation}
  \label{eq:LO2loopdivergence}
  \lim\limits_{q\to\infty}
  \calA^{(\text{S}\leftrightarrow\text{P})}_{LO,2\text{-loop}}\sim 
  (\kv\cdot\vec{K}_{PV})\,q^{1-s_0(\lambda_\text{S})-s_{1}(\lambda_\text{P})}
  \to0
  \;\;,\;\;\Delta^{(\text{S}\leftrightarrow\text{P})}_{LO,2\text{-loop}}
  (\lambda_\text{S},\lambda_\text{P})= 
  1-\textrm{Re}[s_0(\lambda_\text{S})+s_1(\lambda_\text{P})]
\end{equation}
where $\lambda_\text{S}$ ($\lambda_\text{P}$) is the spin-isospin index of the
S-wave (P-wave) PC 3N half-offshell amplitude. The superficial degrees of
divergence are independent of the order of S- and P-waves:
\begin{eqnarray}
  \label{eq:LOdegree2loop}
  \Delta^{(\text{S}\leftrightarrow\text{P})}_{LO,2\text{-loop}}
  (\lambda_\text{S}=\lambda_\text{P}=1)=-1.86\dots&,&
  \Delta^{(\text{S}\leftrightarrow\text{P})}_{LO,2\text{-loop}}
  (\lambda_\text{S}=1,\lambda_\text{P}=-\half)=-0.77\dots\\
  \Delta^{(\text{S}\leftrightarrow\text{P})}_{LO,2\text{-loop}}
  (\lambda_\text{S}=-\half,\lambda_\text{P}=1)=-3.03\dots&;&
  \Delta^{(\text{S}\leftrightarrow\text{P})}_{LO,2\text{-loop}}
  (\lambda_\text{S}=\lambda_\text{P}=-\half)=-2.94\dots
\end{eqnarray}
Since all amplitudes converge at LO, no PV 3NIs are needed for
renormalisability.

\subsection{Structure of PV Three-Nucleon Interactions}
\label{sec:PV3NI}

At LO, we demonstrated that no PV 3NI enters by explicitly showing that no
divergence occurs in LO diagrams with PV 2NIs. The additional suppression by
$Q$ at NLO may come from one more power of loop momentum $q$, and hence may
add one unit to the degrees of divergence in
Eqs.~(\ref{eq:LOdegreeS}/\ref{eq:LOdegreeP}/\ref{eq:LOdegree2loop}).  However,
the number and complexity of possibly divergent diagrams is significantly
larger and arguments based on the superficial degree of divergence become more
elaborate, especially for overlapping divergences. In addition, recall that
the superficial degree of divergence of any diagram provides only an upper
bound.  The actual degree of divergence can be lower when spin-isospin
symmetry and the details of the interactions are taken into account. As this
will be the case, we choose a different method.

In order to demonstrate that there are no PV 3NIs at NLO in the $Nd$ system,
we proceed in three steps: First, construct the spin-isospin-cluster structure
of all PV 3NIs with only one derivative for S-P wave transitions; second,
identify the corresponding structures in all NLO diagrams with PV
2NIs; and third, demonstrate that none of these matches the
spin-isospin-cluster structure of the PV 3NIs, so that a promotion of PV 3NIs
from the simplistic estimate \NXLO{2} to NLO is not required as it cannot
renormalise any potential divergence.

With the iso-doublet nucleon-deuteron system as both in and out state, a PV
3NI has $\Delta I\in\{0;1\}$. A PV interaction with $\Delta I =2$ requires an
$I=3/2$ state. Since the strong interactions considered here do not change
isospin, returning to the $I=1/2$ nucleon-deuteron system would not be
possible without a second, highly suppressed insertion of another PV
interaction. We therefore neglect the $\Delta I=2$ PV 3NI. If an un-physical
$I=3/2$ state is found in both the initial and final state, a $\Delta I=3$ PV
3NI would have to be considered as well. Since the transition amplitude
relates S- and P-waves, the PV 3NI is expected to contain an odd number of
derivatives. In addition, from
Eqs.~(\ref{eq:LOdivergenceS}/\ref{eq:LOdivergenceP}/\ref{eq:LO2loopdivergence}),
the potentially divergent amplitudes are proportional to one power of a
low-energy momentum $p\sim k$. Therefore, a PV 3NI must contain exactly one
derivative. Further building blocks are the nucleon fields $N$, the spin and
iso-spin Pauli matrices and the Levi-Civit\'a symbols $\epsilon^{ijk}$ and
$\epsilon^{ABC}$. Indices must be saturated completely, and only spin and
vector indices can relate. Unitarity, time-reversal symmetry and total angular
momentum conservation must be taken into account, too.

Using \emph{Mathematica} for algebra\"ic manipulations, we find that there is
only one unique iso-scalar structure, with different variants related by Fierz
transformations:
\begin{eqnarray}
  \label{eq:PV3NI-isoscalar}
  \lefteqn{\big(N^\dagger N\big)\,\big(N^\dagger N\big) 
    \,\big(N^\dagger\sigma^i\,\ii\de_i N\big)=
    \big(N^\dagger\sigma^iN\big)\,
    \big(N^\dagger N\big)\,\big(N^\dagger\,\ii\de_i N\big)}\\
  &&=\frac{1}{3}\, \big(N^\dagger \tau_AN\big)\,\big(N^\dagger \tau^AN\big) 
  \,\big(N^\dagger\sigma^i\,\ii\de_i N\big)=
  -\frac{1}{3}\,\big(N^\dagger\sigma_j\tau_AN\big)\,
  \big(N^\dagger\sigma^j\tau^AN\big)\,\big(N^\dagger\sigma^i\,\ii\de_i N\big)
  \nonumber\\
  &&=-\frac{1}{5}\, \big(N^\dagger \sigma_jN\big)\,\big(N^\dagger \sigma^jN\big) 
  \,\big(N^\dagger\sigma^i\,\ii\de_i N\big)
  =\mbox{ etc.} \nonumber
\end{eqnarray} 
In order to construct this interaction in the cluster-decomposition basis,
Eq.~\eqref{eq:clusterdecomp}, note that the components of the \twoS state
couple the auxiliary fields with a nucleon via two forms related by Fierz
identities:
\begin{equation}
  \label{eq:Swave}
  \sigma_i d_t^i N^b = -\left(\tau_A\right)^b{}_c\, d_s^A N^c\;\;,
\end{equation}
where the iso-spinor indices are made explicit. The corresponding
$\twoPone$-wave components contain a spatial derivative whose index is
saturated with another Pauli matrix:
\begin{equation}
  \label{eq:Pwave}
 d_t^i(\vec\sigma\cdot\!\LRd)\sigma_iN^b=
 -d_s^A(\vec\sigma\cdot\!\LRd)\left(\tau_A\right)^b{}_c \,N^c 
\end{equation}
For the \twoS-\twoPone $\Delta I=0$ 3NI, spinor indices are contracted to form
an iso-scalar:
\begin{eqnarray}
  \label{eq:PV3NI-isoscalar-cluster}
  &&\Big[N^\dagger d^{j\dagger}_t\,\sigma_j\Big]_a\delta^a_b
  \Big[d^k_t\,\big(\sigma^i\,\ii\!\!\LRd_i\big)\,\sigma_k\,N\Big]^b+\text{H.c.}
  \\
  &&=-\Big[N^\dagger d^{A\dagger}_s\,\tau_A\Big]_a\delta^a_b
  \Big[d^k_t\,\big(\sigma^i\,\ii\!\!\LRd_i\big)\,\sigma_k\,N\Big]^b+\text{H.c.}
  \nonumber\\ &&
  =\Big[N^\dagger d^{A\dagger}_s\,\tau_A\Big]_a\delta^a_b
  \Big[d^B_s\,\big(\sigma^i\,\ii\!\!\LRd_i\big)\,\tau_B\,N\Big]^b+\text{H.c.}
  \nonumber\\ &&
  =-\frac{9}{2}\big(N^\dagger N\big)\,\big(N^\dagger N\big) 
    \,\big(N^\dagger\sigma^i\,\ii\de_i N\big)\nonumber
\end{eqnarray} 
The last line is the result of yet another Fierz transformation and
establishes the identity between the only possible $\Delta I=0$ PV 3NI with
only one derivative and the cluster-decomposed form of the PV 3NI in the
\twoS-\twoPone channel. That a PV 3NI with only one derivative can be
re-written in this form is not surprising. With no derivative acting on two
combinations containing $N$ and $N^\dagger$ in \eqref{eq:PV3NI-isoscalar},
these two can be re-arranged in a relative S-wave and hence represented by the
S-wave auxiliary fields. The derivative puts the remaining nucleon in a P-wave
relative to the other two particles.

The \twoS-\twoPone $\Delta I=1$ PV 3NI is obtained by inserting $\tau_3$ as
the $\Delta I=1$ operator:
\begin{align}\label{eq:PV3NI-isovec}
 &\Big[N^\dagger d^{j\dagger}_t
   \sigma_j\Big]_a(\tau_3)^a{}_b\Big[d^i_t(\vec \sigma \cdot \ii\!\!\LRd ) \sigma_i
   N\Big]^b+\text{H.c.}\notag\\  
 & =-\Big[N^\dagger d^{A\dagger}_s \tau_A\Big]_a(\tau_3)^a{}_b
 \Big[d^i_t(\vec \sigma \cdot \ii\!\!\LRd ) \sigma_i
   N\Big]^b+\text{H.c.}\notag\\ 
 &=\Big[N^\dagger d^{A\dagger}_s \tau_A\Big]_a(\tau_3)^a{}_b
 \Big[d^B_s(\vec \sigma \cdot \ii\!\!\LRd )
   \tau_B N\Big]^b+\text{H.c.}\notag\\ 
 &= \frac{9}{2} \big(N^\dagger N\big)\,\big(N^\dagger N\big)
    \,\big(N^\dagger\tau_3\sigma^i\,\ii\de_i N\big)=\mbox{ etc.}
\end{align}
As in the iso-singlet case, one can show with the help of algebra\"ic
manipulation software that every PV 3NI with the above-mentioned building
blocks, supplemented by $\tau_3$, is proportional to this structure.  Note
that there is no parity-violating \twoS-\fourPone three-nucleon \emph{contact}
term. The $Nd$ system can of course be in a \fourPone channel.

\begin{figure}[!htb]
  \begin{center}
    \includegraphics*{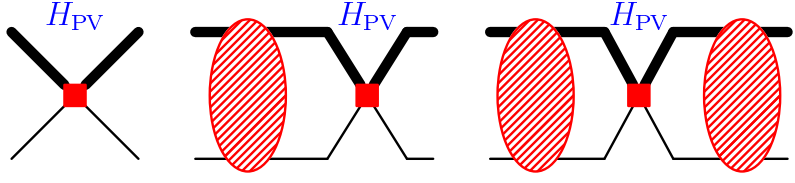}
    \caption{PV 3NI graphs which may absorb infinities.  Crossed contributions
      not displayed.}
    \label{fig:PV3NI-diagrams}
  \end{center}
\end{figure}

The Lagrangean describing these interactions is therefore
\begin{eqnarray}
\label{eq:PV3NILag}
\calL_\text{PV}^\text{3NI}&=&\frac{y^2 M}{3\Lambda^3}\;\bigg[
H_\text{PV}^{(\Delta I=0)}(\Lambda) \;\Big[N^\dagger
d^{j\dagger}_t\,\sigma_j-N^\dagger d^{A\dagger}_s\,\tau_A\Big]
\Big[d^i_t(\vec \sigma \cdot \ii\!\!\LRd ) \sigma_i
N- d^B_s(\vec \sigma \cdot \ii\!\!\LRd )\tau_B N\Big]
\\ &&
+H_\text{PV}^{(\Delta I=1)}(\Lambda)\;\Big[N^\dagger
d^{j\dagger}_t\,\sigma_j-N^\dagger d^{A\dagger}_s\,\tau_A\Big]
\tau^3\Big[d^i_t(\vec \sigma \cdot \ii\!\!\LRd ) \sigma_i
N- d^B_s(\vec \sigma \cdot \ii\!\!\LRd ) \tau_B N\Big]\bigg]+\text{H.c.}
\;.\nonumber 
\end{eqnarray}
The pre-factor was chosen in analogy to that of the PC 3NI in
Eq.~\eqref{eq:PCLag} such that the PV 3NIs $H_\text{PV}^{(\Delta
  I=0,1)}(\Lambda)$ have mass-dimension zero. If PV 3NIs are needed at NLO to cure divergences, $H_\text{PV}$
  may diverge for large $\Lambda$ and enter through the diagrams in
  Fig.~\ref{fig:PV3NI-diagrams}.

Consider now the spin-isospin-cluster structure of the PV 3NIs.
The projector in cluster-configuration space onto the doublet-S wave was
already constructed in Ref.~\cite{effrange1,effrange2} and is supplemented here by that
onto the \wave{2}{P}{\half} channel. Both can also be read off from
(\ref{eq:Swave}/\ref{eq:Pwave}): 
\begin{equation}
  \label{eq:projectors}
  \calP[\wave{2}{S}{\frac{1}{2}}]_{iA}=\frac{1}{\sqrt{3}}\;\begin{pmatrix}
    \sigma_i&0\\0&\tau_A\end{pmatrix}
  \;\;,\;\;
  \calP[\wave{2}{P}{\frac{1}{2}}]_{iA}=\left(\vec{\sigma}\cdot{\ev}\right)
  \frac{1}{\sqrt{3}}\;\begin{pmatrix}
    \sigma_i&0\\0&\tau_A\end{pmatrix}
\;\;,
\end{equation}
where $\ev$ is the unit vector in the direction of the momentum of the
auxiliary field $d_{s/t}$ in the centre-of-mass system. The vector index $i$
and iso-vector index $A$ are contracted with the respective indices of the
auxiliary fields. Since this leaves only free spinor and iso-spinor indices,
the resulting state
\begin{equation}
  \calP[{}^{2}{l}_{\frac{1}{2}}]_{iA}\;{d_t^i\choose d_s^A}\;N
\end{equation}
carries total spin $\half$, and zero or one unit of orbital angular momentum
$l$. The projectors are ortho-normalised as
\begin{equation}
  \frac{1}{4\pi}\int\deint{}{\Omega_e}
  \calP[{}^{2}l_\half]_{iA}\;\calP^\dagger[{}^{2}l^\prime_\half]^{iA}=
  \delta^l_{l^\prime}\;\;,
\end{equation}
where the integration is over the direction of $\vec{e}$. A complete
list of all S- and P-wave projectors of the three-nucleon system is given in
an upcoming publication~\cite{SpinRot}. 

Projecting the PV 3NIs Eq.~\eqref{eq:PV3NILag} onto an incoming \twoS- and
outgoing \twoPone-wave (or vice versa), one finds
\begin{eqnarray}
  \ii\calM\left[\twoS \to \twoPone,p,q\right]_\text{3NI}&=&
    A_\text{3NI}
    \left(H_\text{PV}^{(\Delta I=0)}+\tau^3\,H_\text{PV}^{(\Delta I=1)}\right)
   \begin{pmatrix}1&-1\\-1&1\end{pmatrix},
   \label{eq:PV3NI}
\end{eqnarray}
where the overall factor $A_\text{3NI}$ is a function of
$E,k,p,q,\gamma_{s/t}$ and a scalar in spin-isospin-cluster space. Its exact
form is not needed in the following.  As pointed out in Sect.~\ref{sec:pclag},
a more convenient cluster-space basis in the UV limit is that in which the PC
3N amplitudes are diagonal with spin-isospin parameters $\lambda=1$ or
$\lambda=-\half$. The basis-change \eqref{eq:lincombforWigner}
\begin{eqnarray}\label{eq:PV3NI-Wigner}
  \ii\calM\left[\twoS \to \twoPone,p,q\right]_\text{3NI}^\text{Wigner}&=&
  \half\begin{pmatrix}1&-1\\1&1\end{pmatrix}
  \ii\calM\left[\twoS \to\twoPone,p,q\right]_\text{3NI}
    \begin{pmatrix}1&1\\-1&1\end{pmatrix}\nonumber\\
  &=&A_\text{3NI}\left(H_\text{PV}^{(\Delta
      I=0)}+\tau^3\,H_\text{PV}^{(\Delta I=1)}\right)
  \begin{pmatrix}2&0\\0&0\end{pmatrix}
\end{eqnarray}
shows that both PV 3NIs connect only the 
$\twoS(\lambda=1)$-$\twoPone(\lambda=1)$ components, just as the PC 3NI.
As shown in Sec.~\ref{sec:pclag}, the LO parity-conserving amplitudes are
diagonal in Wigner space. Therefore, all diagrams of
Fig.~\ref{fig:PV3NI-diagrams} have the same structure, i.e.~they all connect
only the $\twoS(\lambda=1)$-$\twoPone(\lambda=1)$ components.

We re-emphasise the most important aspect of this construction: There are two
and only two independent structures with $\Delta I\in\{0;1\}$ available in the
nucleon-deuteron system which contain exactly one derivative, and both are
nonzero \emph{only} in the $\twoS(\lambda=1)$-$\twoPone(\lambda=1)$ channel,
i.e.~only the upper-left entry in the cluster matrix in the Wigner-basis is
non-zero, see Eq.~(\ref{eq:PV3NI-Wigner}).  This implies that any superficial
divergence which does \emph{not} share these quantum numbers cannot be cured
by this PV 3NI. In particular, no divergences can appear at this order in the
\twoS-\fourPone amplitudes or any other channel.

\subsection{Contributions with PV 2NIs at Next-To-Leading Order}
\label{sec:NLOgraphs}

\begin{figure}[!htb]
  \begin{center}
    \includegraphics*{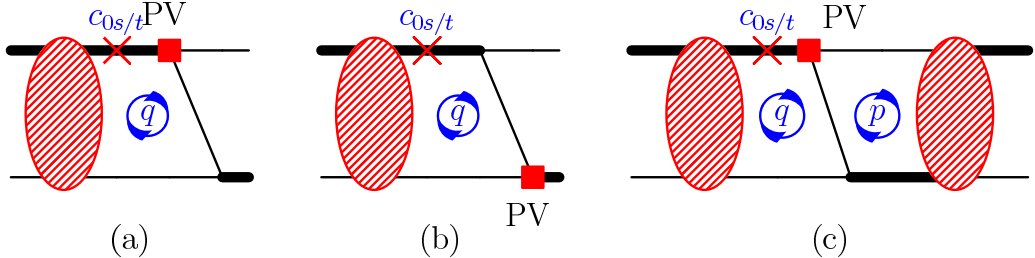}
    \caption{NLO PV diagrams, derived from Figs.~\ref{fig:oneloop} and
      \ref{fig:twoloop} by adding one insertion of the effective-range
      correction to the deuteron propagator. Crossed contributions not
      displayed.}
    \label{fig:NLO}
  \end{center}
\end{figure}

\begin{figure}[!htb]
  \begin{center}
    \includegraphics*{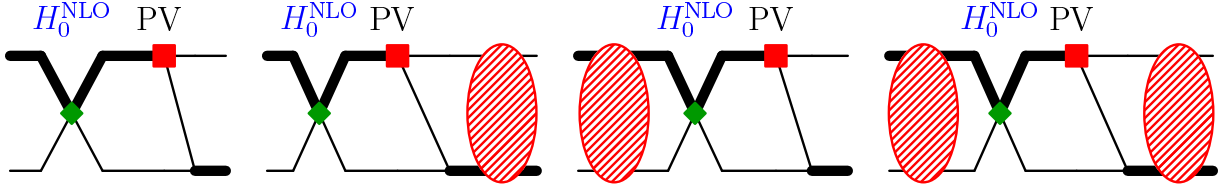}
    \caption{NLO PV diagrams with one insertion of the NLO PC 3NI correction.
      Crossed contributions and those with the PV interaction instead on the
      bottom right not displayed.}
    \label{fig:H0contrib}
  \end{center}
\end{figure}

\begin{figure}[!htb]
  \begin{center}
    \includegraphics*{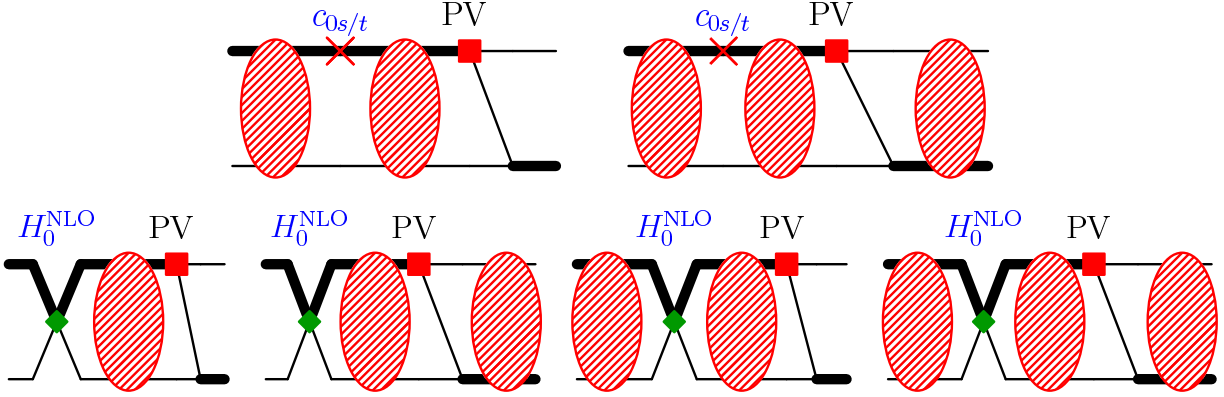}
    \caption{NLO PV diagrams with nucleon-dibaryon rescattering. Crossed
      contributions and those with the PV interaction instead on the bottom
      right not displayed.}
    \label{fig:trivialNLO}
  \end{center}
\end{figure}

In the second step, we write down all NLO diagrams with PV 2NIs.  PV 2NIs
appear only at odd powers of Q.  Thus, higher-order PV corrections only start
contributing at order $\epsilon Q$, i.e.~\NXLO{2}, where simplistic arguments
suggest that PV 3NIs enter as well.  The NLO contributions to PV
nucleon-deuteron scattering therefore stem from the PC corrections presented
in Sec.~\ref{sec:pclag}. The first set, namely effective-range corrections to
Figs.~\ref{fig:oneloop} and \ref{fig:twoloop}, are shown in
Fig.~\ref{fig:NLO}.
The only other PC NLO correction, $H_0^\text{NLO}$, leads to the diagrams of
Fig.~\ref{fig:H0contrib}.
Since this 3NI appears only in the \twoS-wave, the LO amplitude
$t^{(l=0)}_\lambda$ can be attached only on the side of $H_0^\text{NLO}$. 
Finally, a LO amplitude $t^{(l)}_\lambda$ can be
inserted between the PC and PV interactions, see Fig.~\ref{fig:trivialNLO}.
These nucleon-deuteron rescattering contributions are found by convoluting NLO
parity-conserving amplitudes with the LO PV kernels of Fig.~\ref{fig:tree}.

\subsection{No PV Three-Nucleon Interaction at Next-To-Leading Order}
\label{sec:NLOamplitudes}

Finally, we test whether the spin-isospin-cluster structure of these diagrams
matches that of the PV 3NIs, Eq.~\ref{eq:PV3NI-Wigner}.  All contain the same
PV tree-level diagrams of Fig.~\ref{fig:tree}. Using the PC and PV Lagrangeans
(\ref{eq:PCLag}/\ref{eq:PVLag}) and the partial-wave projectors
\eqref{eq:projectors} onto an incoming \twoS- and outgoing \twoPone-wave, the
corresponding structure is:
\begin{equation}
\ii\calM\left[\twoS \to \twoPone,p,q\right]_\text{2NI}=
A^{(a)}_\text{2NI}
  \begin{pmatrix}\calS_1&-\calT\\\calS_1&-\calT\end{pmatrix}+
  A^{(b)}_\text{2NI}
  \begin{pmatrix}\calS_1&\calS_1\\-\calT&-\calT\end{pmatrix}\:.
\label{eq:spinologyNLO}
\end{equation}
Here, $\calS_1=3\daR +2\tau_3\, \deR$ and $\calT=3\dbR +2\tau_3 \,\dcR$ are
combinations of the PV 2NI strengths. The overall factors $A^{(a)}_\text{2NI}$
and $A^{(b)}_\text{2NI}$ are different for the PV 2NI being at the top or
bottom vertex of Fig.~\ref{fig:tree}. Like the factor $A_\text{3NI}$ of
Eq.~\eqref{eq:PV3NI-Wigner}, their dependence on $E,k,p,q,\gamma_{s/t}$ is
irrelevant for the present argument, except for the fact that they may contain
potential divergences. Their exact forms will be provided in the
above-mentioned upcoming article~\cite{SpinRot}. Under the transformation to
the Wigner-basis \eqref{eq:lincombforWigner}:
\begin{eqnarray}
  \label{eq:spinologyNLOWigner}
  \ii\calM\left[\twoS \to \twoPone,p,q\right]^\text{Wigner}_\text{2NI}=
    A^{(a)}_\text{2NI}
  \begin{pmatrix}0&0\\\calS_1+\calT&\calS_1-\calT\end{pmatrix}
  +A^{(b)}_\text{2NI}
  \begin{pmatrix}0&\calS_1+\calT\\0&\calS_1-\calT\end{pmatrix}\:.
\end{eqnarray}
In other words, the structure of the ``tree-level'' diagrams does \emph{not}
contain the only component of the PV 3NIs, i.e.~the
$\twoS(\lambda=1)$-$\twoPone(\lambda=1)$ component (upper-left entry in the
cluster matrix).

As seen in the NLO diagrams of Figs.~\ref{fig:NLO} to~\ref{fig:trivialNLO},
the PV 2NI kernel of~\eqref{eq:spinologyNLO} are convoluted with LO PC
amplitudes and/or NLO corrections in the strong sector. The cluster-space
structure of such graphs is obtained by multiplying the $2\times2$ matrix~\eqref{eq:spinologyNLOWigner} from the left and right with that of:
\begin{enumerate}
\item the LO half- and full-offshell amplitude of
  Eqs.~(\ref{eq:doubletpw}/\ref{eq:doubletpwWigner});
\item the effective-range insertions $c_{0s/t}$ of
  Eqs.~(\ref{eq:PCLag}/\ref{eq:effrangewigner});
\item the NLO parity-conserving 3NI $H_0^\text{NLO}$ of Eq.~\eqref{eq:PCLag}.
\end{enumerate}
However, we demonstrated in Sec.~\ref{sec:pclag} that all these terms are
diagonal in the Wigner-basis since Wigner-symmetry is approximatively realised
in the strong sector. Off-diagonal elements are suppressed by at least one
additional power of $Q$ and hence enter at most at \NXLO{2}.  Any combination
therefore results in a matrix which, at NLO, has a zero in the upper left
corner, $(\lambda=1)\leftrightarrow(\lambda^\prime=1)$:
\begin{equation}
    \begin{pmatrix} 0 & a\\ b & c \end{pmatrix}\;\;,
\end{equation}
with the specific form of the entries $a,b,c$ irrelevant for our argument.
In addition, off-diagonal elements of the PC LO amplitude are suppressed
by $1/q$, making loop integrals more convergent.

As shown in the preceding Section, though, the expressions for the NLO
diagrams with the PV 3NI of Eq.~\eqref{eq:PV3NI-Wigner}
(Fig.~\ref{fig:PV3NI-diagrams}) contain at this order a non-vanishing
entry only in the upper left corner at this order. Any possible divergence at
NLO therefore does not match the cluster-structure of the PV 3NIs.  With the
exception of an anomalously large coefficient, no reason exists to promote the
PV 3NI to either LO or NLO, and it enters not earlier than at \NXLO{2}, in
agreement with the simplistic counting.

Consequently, the NLO PV diagrams can at most contain divergences which are
already renormalised in the parity-conserving sector as in
Fig.~\ref{fig:PCNLO}. This is indeed verified by explicit numerical
computations in an upcoming publication~\cite{SpinRot}.

In summary, we have demonstrated that no PV 3NIs enter in the $Nd$ system at
NLO.


\section{Conclusions}
\label{sec:Conclusions}

We have shown that no parity-violating three-nucleon interaction enters in the
nucleon-deuteron system at leading and next-to-leading order in ``pionless''
Effective Field Theory.  The mass dimensions of PV 3NIs suggest that they
enter at order $\epsilon Q$ (\NXLO{2}).  However, a PV 3NI can \emph{a priori}
be included at lower orders if it is needed as counter-term to absorb
divergences in amplitudes containing PV 2NIs. At LO, $\calO(\epsilon Q^{-1})$,
the superficial degree of divergence of each contribution to the $Nd$
scattering amplitudes is negative, i.e.~all amplitudes converge.  At NLO,
$\calO(\epsilon Q^{0})$, we analysed the structure of the interactions in the
so-called Wigner-basis to show that the quantum numbers of any divergence that
might potentially arise do not match those of the only two PV 3NIs with linear
momentum dependence. We used that Wigner symmetry is approximatively
realised in the UV limit of the strong sector even when effective ranges are
included. Thus, only the five couplings of the LO PV two-nucleon Lagrangean
contribute up to and including NLO on the parity-violating side.

Since PV 3NIs are absent at NLO, the uncertainties of a PV calculation with
\EFTNoPion in this system are $\lesssim10\%$ and therefore competitive with
those of the most ambitious experiments. Extending the analysis to \NXLO{2} is
thus not only unnecessary at present; it would also lead to more free
parameters since additional PV 2NIs enter, and PV 3NIs are expected to
contribute as well.

One might question whether a detailed analysis was indeed necessary to arrive
at a result which is also suggested by a simplistic derivative-counting. It
must however be stressed that the presence of a 3NI already at LO in the
parity-conserving sector can enhance parity-violating 3NI contributions. Given
the experimental difficulties of extracting the coupling constants of the LO
PV 2NI Lagrangean, PV 3NIs at LO or NLO would have significantly compromised
the programme to analyse the strengths of parity-violating two-nucleon
interactions.  Since reliable, model-independent error-assessments are
necessary to interpret the data, investigating the renormalisability of
\EFTNoPion at NLO with parity-violating interactions is therefore crucial.

The investigation here was limited to nucleon-deuteron observables, i.e.~to
the iso-doublet three-nucleon system. The extension to 3N iso-quartet channels
is left to the future. They are experimentally realised only as a sub-system
in processes involving more than $3$ nucleons. in that context, PV 3NIs with
$\Delta I=2,3$ must also be considered. Including electro-magnetic currents
will also be addressed.

Finally, the arguments in this publication are general and prove the
assumption by exclusion. It was the particularly transparent divergence
structure of the PV and PC amplitudes in \EFTNoPion which allowed for
rigorous, analytic arguments without resort to a detailed numerical study of
cut-off dependences in observables.  As a first application of \EFTNoPion in
the three-nucleon sector, we have performed a calculation of the neutron
spin-rotation in deuterium whose results will be available
soon~\cite{SpinRot}.  Schiavilla et al.~\cite{Schiavilla:2008ic} explored this
process in the ``hybrid'' formalism, in which \EFTNoPion PV interactions
are combined with phenomenological wave functions.  In this context, we will
also provide numerical confirmation of the results presented here and the
construction of partial-wave projectors in the $Nd$ system.


\section*{Acknowledgements}
We are deeply grateful to D.~R.~Phillips for providing invaluable suggestions
and encouragement. We also thank J.~Kirscher, M.~Paris, M.~Snow and
R.~P.~Springer for insightful discussions and helpful suggestions, and
W.~Parke for a careful reading of the manuscript.  We are particularly
indebted to the organisers and participants of the INT programme 10-01:
``Simulations and Symmetries'', which also provided financial support.  HWG is
grateful for the kind hospitality of the Nuclear Experiment group of the
Institut Laue-Langevin (Grenoble, France). This work was carried out in part
under National Science Foundation \textsc{Career} award PHY-0645498 and
US-Department of Energy grant DE-FG02-95ER-40907.




\end{document}